\documentclass[11pt,a4paper]{article}
\pdfoutput=1
\usepackage{amsmath}
\usepackage{amssymb}
\usepackage{accents}
\usepackage{authblk}
\usepackage{graphicx}
\usepackage{bm}
\usepackage{color}
\usepackage[utf8]{inputenc} 
\usepackage{cite}
\usepackage{xcolor}
\usepackage{dcolumn}
\usepackage{cancel}
\newcommand{\dpar}[2]{\frac{\partial #1}{\partial #2}}
\newcommand{\dpart}[2]{\tfrac{\partial #1}{\partial #2}}

\title{From diffusion experiments to mean-field theory simulations and back}

\author{M. Di Pietro Mart\'inez and M. Hoyuelos\footnote{hoyuelos@mdp.edu.ar}}

\affil{Instituto de Investigaciones F\'isicas de Mar del Plata (IFIMAR-CONICET) and Departamento de F\'isica, Facultad de Ciencias Exactas y Naturales, Universidad Nacional de Mar del Plata, Funes 3350, 7600 Mar del Plata, Argentina}

\date{\today}

\begin{document}

\maketitle

\begin{abstract}
Using previous experimental data of diffusion in metallic alloys, we obtain real values for an interpolation parameter introduced in a mean-field theory for diffusion with interaction. Values of order $1$ were found as expected, finding relevance for this quantity as a way to better understand the underlying dynamics of diffusion processes. Furthermore, using this theory, we are able to estimate the values of the mean-field potential from experimental data. As a final test, we reobtain, with all this information as an input to our simulations, the diffusion coefficient in the studied metallic alloys.
Therefore, the method provides appropriate transition probabilities to perform Monte Carlo simulations that correctly describe the out of equilibrium behavior.
\end{abstract}

\section{Introduction}

In 1948, Darken \cite{darken} derived two equations that were a major improvement in the understanding of diffusion processes of interacting particles in a wide variety of systems, that found important applications in binary mixtures of metals \cite{sridhar}. He obtained a relationship between the collective diffusion coefficient, $D$ (related to Fick's first law), the single-particle diffusion coefficient, $D^*$ (related to the mean-square displacement), and the activity coefficient $\gamma$ (representing the deviation from ideal behavior of the chemical potential); these three quantities are functions of the concentration $c$. When interactions are negligible, we have that $D=D^*$. The equation that relates the mentioned quantities, for one of the species of the mixture, is given by
\begin{equation}
D = D^{*} \left( 1 + \frac{\partial \ln \gamma}{\partial \ln c} \right)
\label{dark}
\end{equation}
This is a previous step for the derivation of the interdiffusion coefficient $\tilde{D}$ (the diffusivity in the reference frame where volume flux is zero) in terms of the diffusivities of the two species, $D_A$ and $D_B$, in a binary mixture
\begin{equation}
\tilde{D} = x D_A + (1-x)D_B
\label{eqdark2}
\end{equation}
where $x$ is the mole fraction of species $A$; the label ``Darken equation'' can be found in the literature applied to Eq.\ \eqref{dark} or \eqref{eqdark2}. Besides metallic mixtures, Darken equations were applied to ceramics \cite{cooper}, polymers \cite{kramer} and diffusion in zeolite \cite{skoul}; see also Ref.~\cite{sridhar}.

A mean-field approach to diffusion with interaction was introduced in Ref.~\cite{suarez}. The system is divided in cells and transition probabilities between neighboring cells depend on a mean-field potential $V$, that represents the potential for one particle generated by the presence of all the others. The interaction is local, i.e., its range is small enough so that interactions among particles that are in different cells can be neglected. An important ingredient of the approach is an interpolation parameter $\theta$ that determines if the transition probability, of a particle that jumps from a position to another, depends on the mean-field potential at the origin position, the target position, or a combination of both. Specifically, the transition probability $W_{i,i+1}$ for a particle to jump from cell $i$ to a neighboring cell $i+1$, in the absence of an external potential, is
\begin{equation}
W_{i,i+1} = P \exp \left[ -\frac{\beta}{2} \left( \theta  (V_{i+1} + V_{i}) + \Delta V \right) \right],
\label{eq:transition}
\end{equation}
where $P$ is the rate of jump attempts, $\beta$ is the Boltzmann constant and $\Delta V = V_{i+1} - V_i$. If $\theta = -1$, it can be seen that $W$ depends on the potential $V_{i}$ in the origin cell $i$, while for $\theta = 1$ it depends on the potential $V_{i+1}$ in the destination cell $i+1$. If $\theta=0$, then $W_{i,i+1}$ depends on the difference $\Delta V$; this is the transition probability that is used in Monte Carlo simulation with Glauber algorithm \cite{janke}, that converges to the correct equilibrium solution but does not guarantee a correct representation of the transient evolution. The role of parameter $\theta$ is to correct the out of equilibrium behavior without modifying the equilibrium solution; it can be shown that the diffusion and mobility coefficients depend on $\theta$, but not the equilibrium solution, see Ref.~\cite{suarez}. Measuring $\theta$ will allow to understand the underlying dynamics of diffusion processes with a simple and minimal model. For example, values of $\theta$ close to zero would validate the out of equilibrium behavior of a Monte Carlo simulation with Glauber algorithm.

In Ref.~\cite{mdp}, we analyzed the case when the diffusion process is described by the concentration of only one species (appropriate for solute-solvent systems or surface diffusion), obtained a direct relationship between the mean-field potential and the activity coefficient and derived Eq.~\eqref{dark}. Here we extend these results to binary mixtures composed by two atomic species that have the same atomic volume, the situation in which Darken equations were originally derived. Besides deriving Eq.~\eqref{dark}, we have also a relationship between the diffusion coefficient and parameter $\theta$. We used experimental results, previously obtained by other authors, of the diffusion coefficient against molar fraction in order to obtain $\theta$ for different metallic binary mixtures.
We obtained values of $\theta$ of order 1, in accordance with the proposed interpretation as an interpolation parameter.

In Sec.~\ref{sec:model} we present the mathematical description of the model for diffusion in binary mixtures, obtain the relationship between mean-field potential and activity coefficient and derive Eq.~\eqref{dark} for constant total concentration.
In Sec.~\ref{sec:param} we obtain values of $\theta$ from experimental data of different metallic mixtures. In Sec.~\ref{sec:expsim} we present results of diffusion coefficient from numerical simulations; we use values of $\theta$ obtained in the previous section for the transition probabilities and verify that the diffusion coefficient matches experimental values.

\section{Model}
\label{sec:model}

We consider a binary mixture of components $A$ and $B$, with molar concentrations $c_A$ and $c_B$, diffusion coefficients $D_A$ and $D_B$, and activity coefficients $\gamma_A$ and $\gamma_B$. We do the calculations for species $A$; they are equivalent for both species. The system is unidimensional, with $L$ cells of length $a$, but the results can be easily extended to higher dimensions (as long as the mean-square displacement is proportional to time, i.e., normal diffusion). In the cell identified with index $i$ there is a number of particles $n_{i}$ of species $A$, which is related to the molar concentration by $c_{i} = n_{i}/(N_{A} a)$, where $N_{A}$ is Avogadro's constant.

For transition probabilities \eqref{eq:transition}, we can obtain the following expression for the mole current in the continuous limit (see Appendix A):
\begin{equation}
J_A = - D_{A0}\, e^{-\beta \theta V_A}\left(\beta c_A \dpar{V_A}{z} + \dpar{c_A}{z} \right)
\label{corrJ}
\end{equation}
where $D_{A0}$ is the free diffusion coefficient for particles $A$, and we use $z$ for the space coordinate. In the limit of small concentration we have that $V_A=0$ and $J_A = - D_{A0} \dpar{c_A}{z}$.

The transition probabilities can also be used to calculate the mean-square displacement and obtain the single-particle or self-diffusion coefficient (see Appendix B):
\begin{equation}
D_A^* = D_{A0} \, e^{-\beta \theta V_A}
\end{equation}

The next step is to obtain the relationship between mean-field potential $V_A$ and activity coefficient $\gamma_A$.

\subsection{Mean-field potential and activity coefficient}

The chemical potential for species $A$ is
\begin{equation}
\mu_A = \mu_A^* + RT \ln (\gamma_A x)
\label{chempotx}
\end{equation}
where $\mu_A^*$ is the chemical potential for pure $A$, and $x = c_A/c_T$ is the mole fraction of species $A$, with $c_T=c_A+c_B$. In principle, we assume that the total concentration $c_T$ is not constant. If $x=1$, then $c_A=c_{A0}$, $c_B=0$, $\gamma_A=1$ and $\gamma_B = \gamma_{B0}$. If $x=0$, then $c_B=c_{B0}$, $c_A=0$, $\gamma_B=1$ and and $\gamma_A = \gamma_{A0}$. The simplifying assumption $c_T \simeq c_{A0} \simeq c_{B0}$ will be analyzed at the end of this section.

Let us compare the expressions for $c_A$ obtained from the chemical potential \eqref{chempotx}
\begin{equation}
c_A = c_T\, e^{-\beta(\tilde{\mu}_A^* - \tilde{\mu}_A + k_B T \ln \gamma_A)}
\end{equation}
where the tilde in $\tilde{\mu}_A$ indicates that the chemical potential is per particle (instead of mole), and from the equilibrium condition $J_A=0$:
\begin{equation}
c_A = c_{A0}'\, e^{-\beta (V_A - \tilde{\mu}_A)}.
\label{cequil}
\end{equation}
where $c_{A0}'$ is a reference concentration. It is easy to verify that \eqref{cequil} is an equilibrium solution by replacing in \eqref{corrJ}. We obtain
\begin{equation}
V_A - k_B T \ln c_{A0}' = \tilde{\mu}_A^* + k_B T \ln \gamma_A - k_B T \ln c_T.
\label{aux1}
\end{equation}
In the limit $x\rightarrow 0$, we have $\gamma_A \rightarrow \gamma_{A0}$, $V_A\rightarrow 0$ and $c_T \rightarrow c_{B0}$:
\begin{equation}
- k_B T \ln c_{A0}' = \tilde{\mu}_A^* + k_B T \ln \gamma_{A0} - k_B T \ln c_{B0},
\label{aux2}
\end{equation}
a relation that holds for any concentration. Replacing \eqref{aux2} in \eqref{aux1}, we obtain
\begin{equation}
\beta V_A = \ln \frac{\gamma_A\, c_{B0}}{\gamma_{A0}\, c_T}
\label{va1}
\end{equation}

The previous calculations are similar to those presented in Ref.~\cite{mdp} for solute-solvent systems where, for comparison, the mean-field potential for the solute is $\beta V=\ln \gamma$. One difference comes from the definitions of the chemical potential and activity coefficient in each case. In solute-solvent systems, the activity coefficient is equal to 1 in the limit of small concentration, while for a binary mixture, it is equal to 1 for $x=1$; see, e.g., Refs.~\cite[ch.~7]{devoe} or \cite[ch.~5]{atkins}. In any case, the mean-field potential tends to zero in the limit of small concentration.

Eq.~\eqref{va1} was obtained in equilibrium conditions, with fixed values of temperature and chemical potential. We have to justify its application in out of equilibrium situations. This can be done using a standard approximation in classical irreversible thermodynamics: local thermal equilibrium. This approximation holds for smooth variations between neighboring cells; the cell length $a$ should be much smaller than the characteristic length of the concentration spatial variations (a necessary condition also for the continuous limit applied in Appendix A). Once this condition is fulfilled, each cell can be described as if it were in equilibrium, although the whole system is not. For the limitations of classical irreversible thermodynamics and local thermal equilibrium, see \cite[Sec.\ 2.7]{lebon}. Despite its limitations, classical irreversible thermodynamics ``has been very useful in dealing with a wide variety of practical problems'' \cite[p. 37]{lebon}.

\subsection{Constant total concentration}

We have the most simple situation when the total concentration $c_T$ can be assumed constant: $c_T = c_A + c_B = c_{A0} = c_{B0}$; this is the assumption used by Darken in his classical paper \cite{darken}. The relation between mean-field potential and activity coefficient reduces to
\begin{equation}
\beta V_{A} = \ln(\gamma_{A}/\gamma_{A0}).
\label{eq:betaV}
\end{equation}
The current of species $A$, Eq.~\eqref{corrJ}, becomes
\begin{equation}
J_A = -D_A^*\left( 1 + \dpar{\ln \gamma_A}{\ln x} \right) \dpar{c_A}{z}
\label{eq:jac}
\end{equation}
with
\begin{equation}
D_A^*/D_{A0} = (\gamma_A/\gamma_{A0})^{-\theta}.
\label{eq:difcol}
\end{equation}
From this last equation we have that parameter $\theta$ is
\begin{equation}
\theta = - \frac{\ln (D_A^*/D_{A0})}{\ln (\gamma_A/\gamma_{A0})}.
\label{eq:theta}
\end{equation}
Therefore, if we have experimental values of $D_A^*$ and $\gamma_A$ for different values of the molar fraction $x$, we can obtain $\theta$ as a function of $x$.

Then, from \eqref{eq:jac}, the collective diffusion coefficient is
\begin{equation}
D_A = D_A^*\left( 1 + \dpar{\ln \gamma_A}{\ln x} \right).
\end{equation}
We have shown that, starting from the transition probabilities defined in terms of the mean-field potential, Eq.~\eqref{dark}, that relates both diffusion coefficients with the activity coefficient, can be derived.

In the limit of low concentration, $x\rightarrow 0$, both diffusion coefficients are equal, since $\dpar{\ln \gamma_A}{\ln x} = \frac{x}{\gamma_A} \dpar{\gamma_A}{x}\rightarrow 0$; they take the value of the free diffusion coefficient $D_{A0}$.

There is a maximal amount of particles per cell $n_{max}$ and it is related to the total concentration by $c_{T} = n_{max}/(N_{A} a)$. Then, the molar fraction $x_{i}$ on a cell is equal to $n_{i}/n_{max}$, a relationship that holds for constant total concentration.

In order to perform diffusion simulations with our model, it is necessary to know $\theta$ and $V$ as a function of the concentration.
The interpolation parameter $\theta$ can be calculated from the experimental data of the activity or activity coefficient and the self-diffusion or diffusivity coefficient using Eq.~\eqref{eq:theta}, while $V$ can be obtained using Eq.~\eqref{eq:betaV}.


\section{Calculation of mean-field parameters from experimental data}
\label{sec:param}
So, is it possible to calculate the parameter $\theta$ in real systems? If so, what values does it take? First, we searched in the literature for mixtures where Darken equation \eqref{dark} is valid. This is important to highlight, as there are many mixtures where Darken relations are not fulfilled \cite{dagostino11, dagostino13}.
Using experimental data from the following references, we calculated $\theta$ and the mean-field potential $V$ as explained in the previous section.
We present results for different metals diffusing in their respective alloys:
Al-Cu at $T=1500$K \cite{cheng, landolt-binarysys1-AlCu},
Au-Ag at $T=1167$K \cite{mead, landolt-binarysys1-AgAu},
and Au-Ni at $T=1173$K \cite{reynolds, seigel, landolt-binarysys5-AuNi, vandal}.

Only in the case of Al-Cu alloy, the diffusion-coefficient data comes from molecular-dynamic (MD) simulations, instead of experiments. As far as we know, there is no experimental data for the diffusion coefficients of this molten mixture that spans the whole range of composition.
But in this sense, using MD data also permits us not to worry about the convection (fluid flow) effect that is of mayor concern in experiments \cite{cheng,lee}, as it is not considered in MD simulations either.

\begin{figure}
\centering
\includegraphics[width=0.7\textwidth]{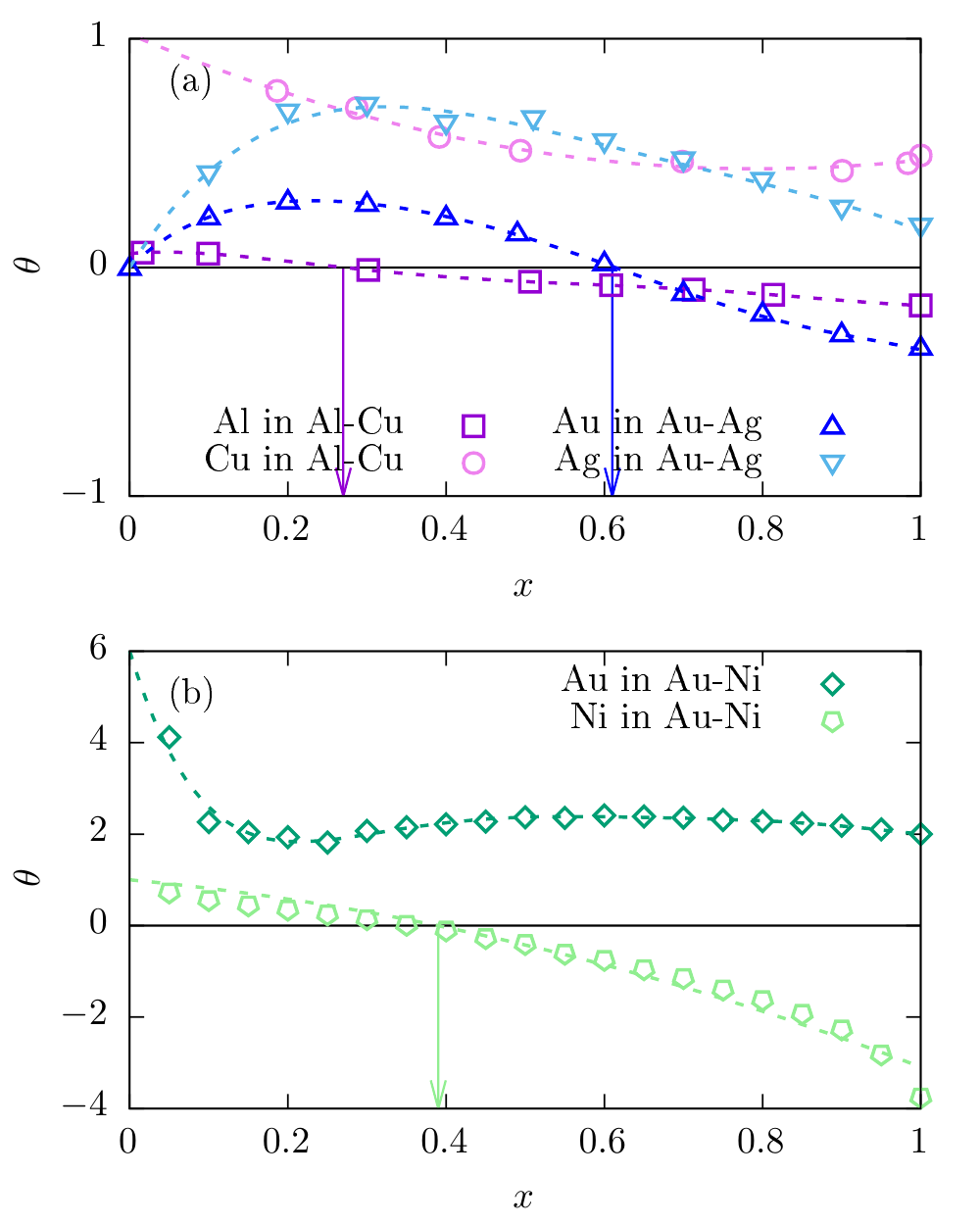}
	\caption{Parameter $\theta$ that characterizes the diffusion process of Al in Al-Cu at $T=1500$K (squares), Cu in Al-Cu at $T=1500$K (circles), Au in Au-Ag at $T=1167$K (up-triangles), Ag in Au-Ag at $T=1167$K (down-triangles), Au in Au-Ni at $T=1173$K (romboids), and Ni in Au-Ni at $T=1173$K (pentagons) as a function of their respective molar fraction $x$ ($x_{\mbox{Al}}$ for Al in Al-Cu, etc.). Dashed lines correspond to polynomial fits. Parameter $\theta$ was calculated using Eq.~\eqref{eq:theta} and data from Refs.~\cite{cheng,mead,reynolds,seigel,landolt-binarysys1-AlCu,landolt-binarysys1-AgAu,landolt-binarysys5-AuNi,vandal}. Additionally, compositions for which $\theta=0$ are marked.}
	\label{fig:theta}
\end{figure}

In Fig.~\ref{fig:theta}, we show the dependence of parameter $\theta$ on molar fraction $x$. Although the curves behave differently, it is important to note that they are all of the same order of magnitude, as well as the fact that $\theta\neq 0$ for most mixtures and compositions, that means, as we mentioned before, that simulations with, for example, the Glauber algorithm would not be the more suitable choice for representing the diffusive dynamics in these mixtures.
In the case of Gold and Nickel diffusing in Au-Ni, values of $\theta$ vary from $-4$ to $6$, while in the others they vary between $-1$ and $1$. This is consistent with the interpretation of $\theta$, introduced in Ref.~\cite{suarez}, as an interpolation parameter.
It is also worth noting that $\theta$ is not constant but it depends on the concentration, which will be reflected in the diffusion coefficient behavior.

\begin{figure}
\centering
\includegraphics[width=0.7\textwidth]{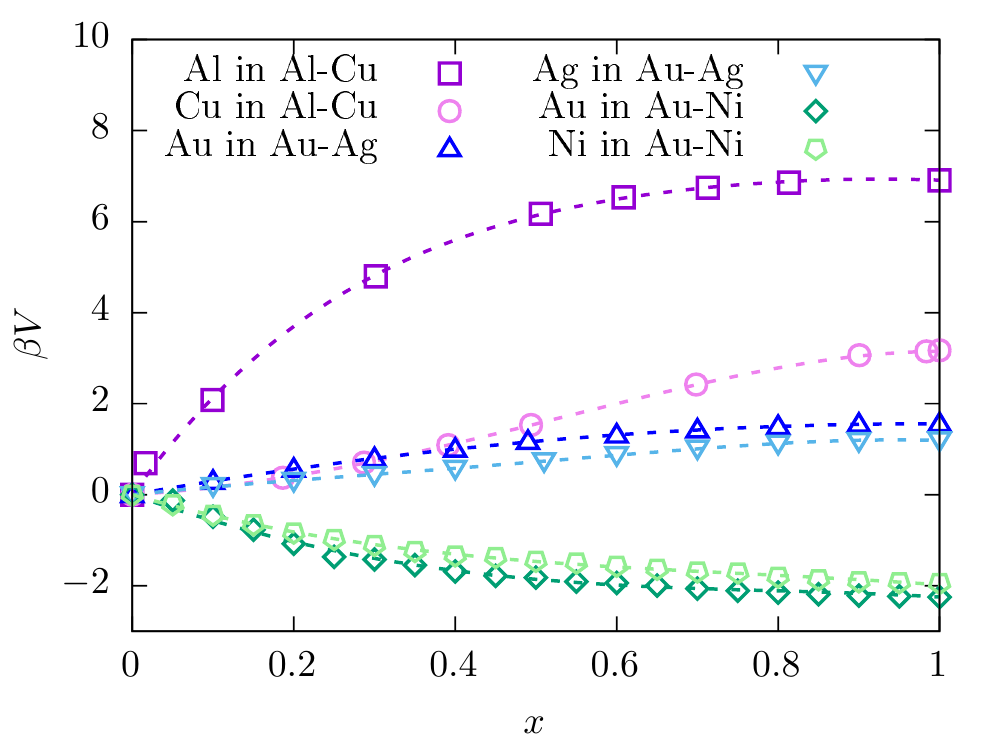}
	\caption{Mean-field potential $\beta V$ of Al in Al-Cu at $T=1500$K (squares), Cu in Al-Cu at $T=1500$K (circles), Au in Au-Ag at $T=1167$K (up-triangles), Ag in Au-Ag at $T=1167$K (down-triangles), Au in Au-Ni at $T=1173$K (romboids), and Ni in Au-Ni at $T=1173$K (pentagons) as a function of their respective molar fraction $x$ ($x_{\mbox{Al}}$ for Al in Al-Cu, etc.). Dashed lines correspond to polynomial fits. Points were calculated using Eq.~\eqref{eq:betaV} and data from Refs.~\cite{cheng,mead,reynolds,seigel,landolt-binarysys1-AlCu,landolt-binarysys1-AgAu,landolt-binarysys5-AuNi,vandal}.}
	\label{fig:betav}
\end{figure}

Fig.~\ref{fig:betav} shows mean-field potential dependence on $x$. These three mixtures are interesting as prototypical examples as they have different behaviors. It can be seen from this figure that the Au-Ni alloy has an attractive potential while Al-Cu and Au-Ag have a repulsive potential, one more pronounced than the other.

In these figures, dashed lines correspond to polynomial fits. In order to use this data in simulations, it is necessary to describe the behaviors with a function, thus the fitting. We get then the functions $\theta(x)$ and $V(x)$ for each mixture. See Appendix C for specific polynomial fits found for each set of data.

\section{Simulation of the experiment}
\label{sec:expsim}
There are a number of methods to measure diffusion coefficients.
Methods based on Fick's first law are known as \emph{direct methods} \cite{Mehrer1990}. These experiments usually consist of a sample, in the shape of a cylindrical tube, where the concentration at the ends is fixed. If these concentrations are different, a measurable diffusion flux will appear and, using Fick's first law, it is possible to calculate the collective diffusion coefficient $D$,
\begin{equation}
 J_n = - D \frac{\partial n }{\partial z}.
 \label{eq:curr}
\end{equation}
where $J_n$ is the particle current and $n$ is the average particle number per cell of length $a$. The diffusion is intended to happen only in the longitudinal direction of the cylinder, as it was a unidimensional system. Also, these experiments occur over distances that are large compared to the interatomic distance, which is what we considered in our model.

\begin{figure}
\centering
\includegraphics[width=0.7\textwidth]{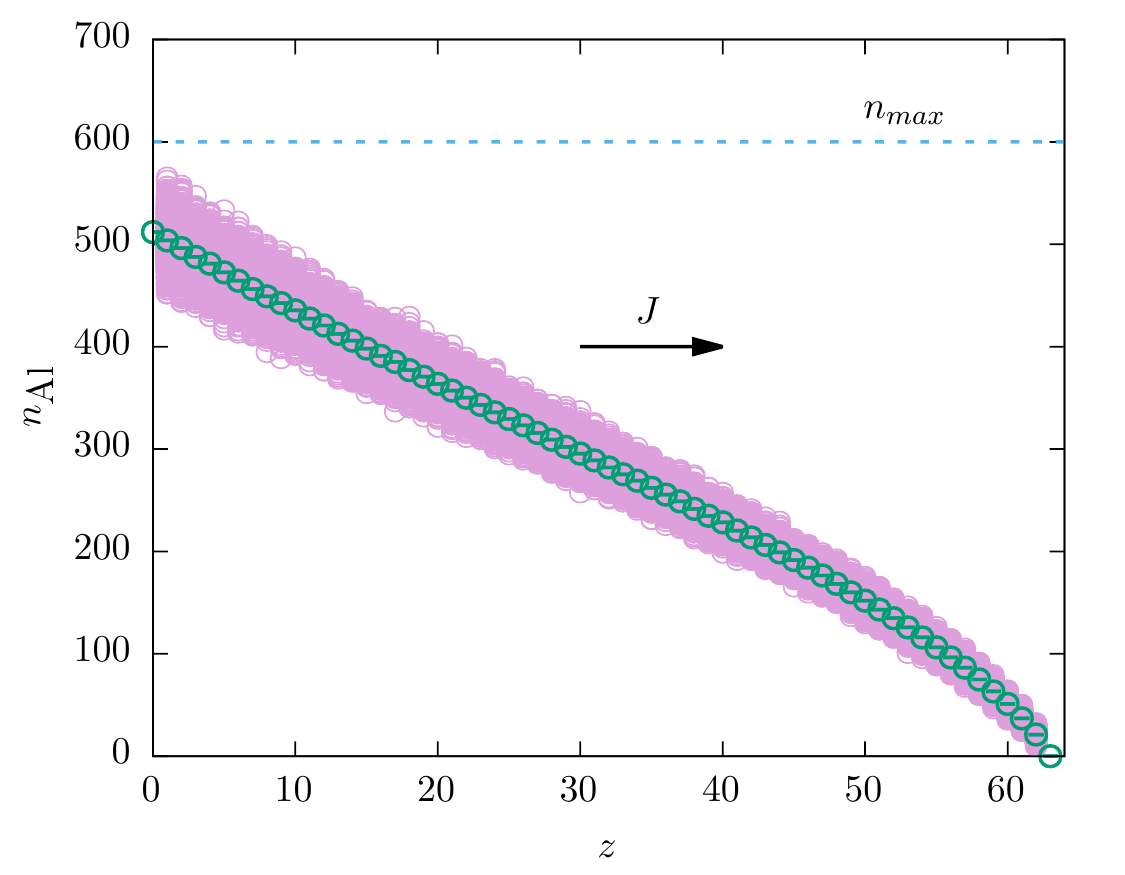}
	\caption{Concentration profile from simulations for Al in Al-Cu. Boundaries' concentrations are fixed so that a steady current $J$ is generated and a diffusive concentration profile appears. Note that $n_{max}$ must be set above the concentration fluctuations so $x$ does not surpass $1$ at any time, therefore it is not possible to reach in average maximal concentration ($\langle x\rangle=1$).}
	\label{fig:dxdz}
\end{figure}

In order to simulate the experiments, let us consider the system introduced in Sec.~\ref{sec:model}.
We fixed the concentration in the first cell to $n_{0} \lesssim n_{max}$ and the concentration in the last cell to zero to mimic the experiment.
After a relaxation time, a steady state is reached, and so it is possible to measure a steady current $J_n$ and a diffusive profile $n(x)$, just like in the experiments.
Fig.~\ref{fig:dxdz} illustrates the explained set up for Aluminum diffusing in Al-Cu alloy.
Note that $n_{max}$ must be set above the concentration fluctuations so $x_i$ does not surpass $1$ at any time.
Remember that $n_{max}$ is the maximum value of the average number of particles in a cell, corresponding to $x_i=1$; it is a constant parameter, while $n_i$ has fluctuations.
The reason for this choice is that, to evaluate transition probabilities, we need the values of $\theta(x)$ and $V(x)$, and they are defined in the range of $x$ between $0$ and $1$. An extrapolation of the fitting curves of $\theta$ and $V$ for $x>1$ is out of the range of available data and would be trustworthy only very close to $x=1$. Therefore, we decided not to cross that limit in the simulations.
Hence, it is not possible to reach, in average, maximal concentration, $\langle x\rangle=1$.
Although results closer to $x=1$ can be obtained by reducing enough the size of fluctuations, \emph{i.e.} increasing appropriately the number of particles in the simulations.

\begin{figure}
\centering
\includegraphics[width=0.8\textwidth]{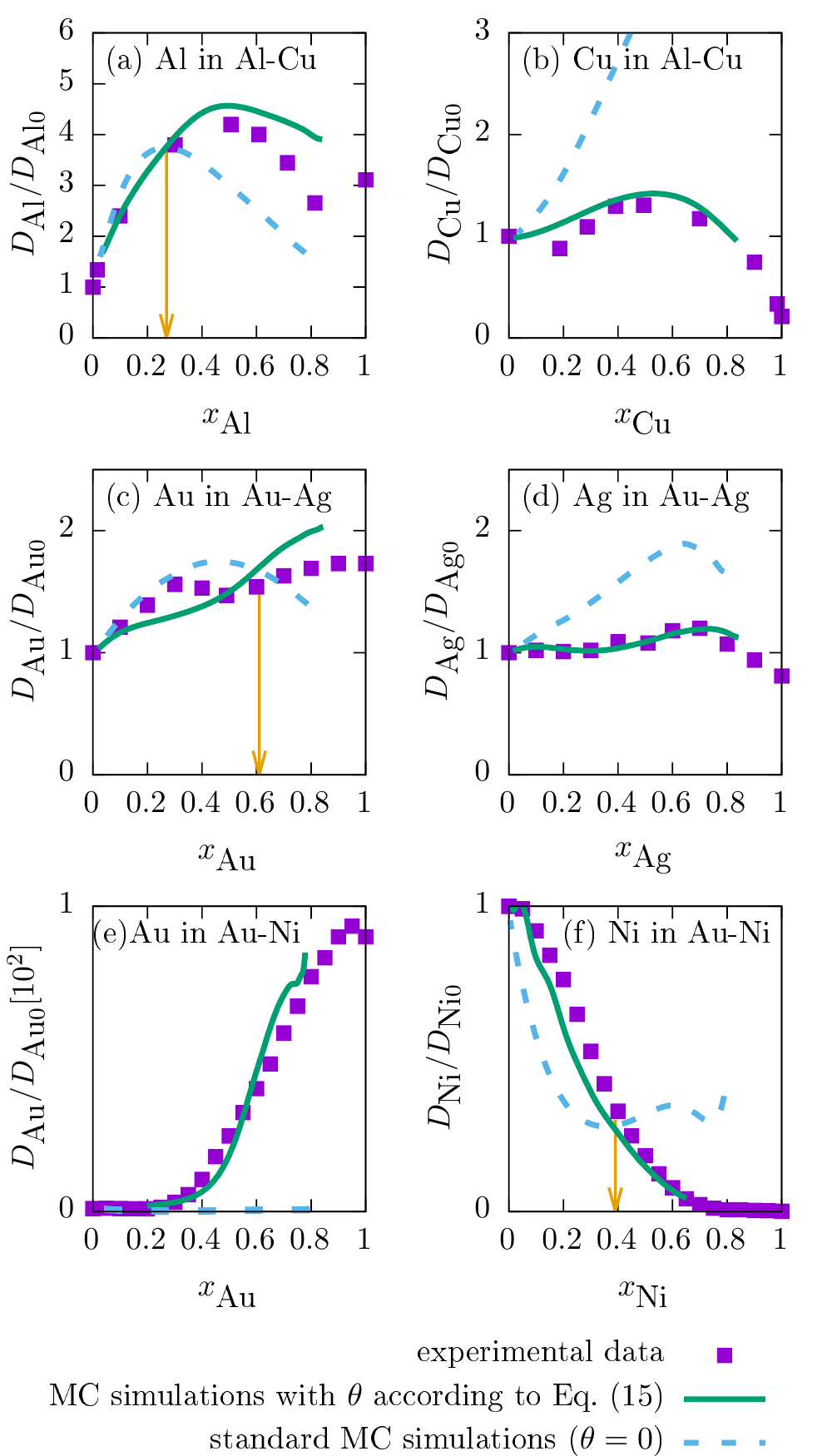}
	\caption{Relative Fickian diffusivity, $D/D_0$, as a function of the molar fraction for different binary mixtures: (a) Al in Al-Cu at $T=1500$K, (b) Cu in Al-Cu at $T=1500$K, (c) Au in Au-Ag at $T=1167$K, (d) Ag in Au-Ag at $T=1167$K, (e) Au in Au-Ni at $T=1173$K and (f) Ni in Au-Ni at $T=1173$K. Data from Refs.~\cite{cheng,mead,reynolds} (squares), simulation of our model using Eqs.~\eqref{eq:betaV} and \eqref{eq:theta} (continuous line) and standard MC simulations (\emph{i.e.} with $\theta=0$) (dashed line) are shown. Additionally, compositions for which $\theta=0$ experimentally are marked with arrows.}
	\label{fig:dif}
\end{figure}

We perform kinetic Monte Carlo simulations over the described system.
The results we show in this paper are for a system of $L=64$, $n_{max}=600$ and $n_{0}=512$ or $n_{0}=480$.
The steady state was found after $10^3$ MC steps and we averaged over $10^5$ steps approximately.
Using Ec.~\eqref{eq:curr}, we obtained from the simulations the diffusion coefficient of each mixture as a function of the molar fraction $x$.
In Fig.~\ref{fig:dif}, we show the results from the simulations and we compare them with the literature data from Refs.~\cite{cheng,mead,reynolds}.

In general, the simulations agree qualitatively with the data.
A motivating good match was found for Aluminum and Cooper diffusing in Al-Cu alloy, for Silver diffusing in Au-Ag and for Nickel diffusing in Au-Ni.
In the case of Gold diffusing in Au-Ag and Gold diffusing in Au-Ni, the simulations fail to reproduce exactly the functional shape.
We found that having better and more experimental data is mandatory in order to get a fit for $V(x)$ and $\theta(x)$ more representative of the physics, since interpolating may generate spurious behavior.

In Fig.~\ref{fig:dif}, we also show simulations using standard MC to contrast with our model.
Here, we use the term ``standard MC'' for simulations with transition probabilities determined only by the energy change, $\Delta V$. Glauber or Metropolis algorithms are generally used to obtain probabilities for fixed time step simulations in this case. We use, instead, kinetic Monte Carlo (stochastic time step). Then, we now consider simulations with transition probabilities given by \eqref{eq:transition} with $\theta=0$.
It can be seen that standard MC results deviate significantly from the data, with some exceptions.
As we mentioned before, when the interpolation parameter is zero, standard MC would be able to represent the dynamics accurately, in this case, the Fickian diffusivity. The results are consistent with this statement as, even though most of the curve does not match, points corresponding to $\theta=0$ do. We have marked these points in Fig.~\ref{fig:theta} and \ref{fig:dif} for an easier comparison.
The deviations grow bigger as $\theta$ departs from zero.

In summary, we found the results encouraging as they validate the relations found \emph{via} the mean-field approach theory.

\section{Discussion and Conclusions}
A mean-field theory of diffusion with interaction was applied to binary mixtures. The main parameters of the theory are the mean-field potential $V_A$ and the interpolation parameter $\theta$. Relationships between these parameters and measurable observables in an experimental set up were presented.

In Sec.~\ref{sec:model}, we showed that the mean-field potential (of one species) is directly related with its corresponding activity coefficient, see Eq. \eqref{eq:betaV}. On the other side, parameter $\theta$ is related to the activity and the diffusion coefficient, see Eq.\ \eqref{eq:theta}.

In Sec.~\ref{sec:param}, we calculated the values for $\theta$ and the mean-field potential as a function of the molar concentration using data of metallic alloys found in the literature: Al-Cu, Au-Ag and Au-Ni. One of the main results of the present work is that values of order of magnitude 1 were obtained for $\theta$. This result is in accordance with the interpretation of $\theta$ as an interpolation parameter that determines if the transition probability between neighboring cells depends on the potential in the origin position, the target position, or a combination of both. The values of $\theta$ were obtained assuming a constant total concentration; a further refinement of the theory is to consider that the total concentration is not constant, this would introduce small corrections to $\theta$ that do not modify the qualitative behavior. There is another assumption that may introduce changes in $\theta$ without modifying its order of magnitude: a possible dependence of the activation energy on molar fraction is not explicitly included. If there is any variation of the activation energy with $x$, we consider that it is included in $\theta$. This is an additional source of variation of $\theta$ that we will analyze more deeply in a future work.

Finally, in Sec.~\ref{sec:expsim}, we used the experimental data of $\theta$ and $V$ as an input for our simulations and we were able to reproduce the experimental behavior of the diffusion coefficient of the studied alloys. Nevertheless, the agreement between numerical and experimental results (Fig.\ \ref{fig:dif}) is not perfect. Some small discrepancies are the consequence of an imperfect fitting of the experimental data used to obtain polynomials in $x$ for $\theta$ and $V$ (see Appendix C).

This is a simple, or minimal, model to describe diffusion with interactions using Monte Carlo simulations. Despite its simplicity, it is able to reproduce the non trivial behavior of the diffusion coefficient against molar fraction (see Fig. \ref{fig:dif}). On the other hand, due to its simplicity it has several limitations. Since the mean field approach reproduces the Darken equation \eqref{dark}, it has, of course, the same limitations of the Darken theory. For diffusion mediated by vacancies in a binary mixture, it is assumed that the vacancy concentration is in thermal equilibrium. If it is not, a vacancy flux causes a vacancy-wind effect that introduces a correction factor in the interdiffusion coefficient, denoted as the Manning factor, see, e.g., \cite[Sec.\ 10.4]{Mehrer2007}.
Detailed molecular dynamics simulations of diffusion in asymmetric mixed plasma show, on the other hand, that the cross-correlation in velocities of different species has an important influence on the interdiffusion coefficient that produces a discrepancy with the Darken equation \cite{haxhimali}.

Monte Carlo simulations of diffusion processes with transition probabilities given, for example, by Glauber or Metropolis algorithms guarantee evolution to equilibrium, but do not guarantee a correct description of the out of equilibrium behavior. The motivation to introduce parameter $\theta$ in the transition probabilities \eqref{eq:transition} is to perform simulations that do not only evolve to equilibrium, but also correctly describe the out of equilibrium behavior. With the appropriate form of $\theta$, that can be obtained from experimental data of the self-diffusion coefficient, we can perform numerical simulations with the correct out of equilibrium behavior. This was verified, as mentioned before, with the comparison between numerical and experimental values of the diffusion coefficient using data of three different binary mixtures, in which the diffusion coefficient has a non trivial dependence on the concentration.

\section*{Appendix A}

We present here the derivation of Eq.~\eqref{corrJ} from the transition probabilities \eqref{eq:transition}. The particle current between cells $i$ and $i+1$ is
\begin{equation}
J_n = n_i W_{i,i+1} - n_{i+1}W_{i+1,i}.
\end{equation}
The expression holds for both species, $A$ and $B$, and we do not specify which one with another subindex in order to simplify the notation of this appendix. Using \eqref{eq:transition} we have
\begin{align*}
J_n =&\ n_i P e^{-\beta\left[ (\theta + 1) V_{i+1} + (\theta-1)V_i \right]/2} \\
& - n_{i+1} P e^{-\beta\left[ (\theta + 1)V_{i} + (\theta-1) V_{i+1} \right]/2}
\end{align*}
In the continuous limit we replace the number of particles $n_i$ by a function $n$ of the space coordinate $z=a\,i$, and the mean-field potential $V_i$ by $V$, a function of $n$. Average over configurations is taken and correlations in non-linear terms are neglected (Ginzburg criterion). We replace $n_{i+1} \rightarrow n + \frac{\partial n}{\partial z} a$ and $V_{i+1} \rightarrow V + \frac{d V}{d z} a$. Then, we have
\begin{align*}
J_n & = P e^{-\beta\theta V} \left[ n \exp\left(-\beta \frac{(\theta+1)}{2}\dpar{V}{z}a \right) \right. \\
& - \left.\left(n+\frac{\partial n}{\partial z}a \right)\,\exp\left(-\beta \frac{(\theta-1)}{2}\dpar{V}{z}a \right)\right].
\end{align*}
Assuming smooth variations of the concentration, the previous expression can be approximated, up to order $a$, by
\begin{align*}
J_n& = P e^{-\beta\theta V} \left[ n \left(1- \beta \frac{(\theta+1)}{2}\dpar{V}{z} a \right) \right. \\
& - \left.\left(n+\frac{\partial n}{\partial z} a\right)\,\left(1- \beta \frac{(\theta-1)}{2}\dpar{V}{z} a\right)\right]
\\
&= -P a\, e^{-\beta\theta V} \left(\beta n \dpar{V}{z} + \frac{\partial n}{\partial z}\right).
\end{align*}
Using that the mole current is $J=J_n/N_A$ and the mole concentration is $c=n/(aN_A)$, we arrive at the result of Eq.~\eqref{corrJ}:
\begin{equation}
J = D_0\, e^{-\beta\theta V} \left(\beta c \dpar{V}{z} + \frac{\partial c}{\partial z}\right)
\end{equation}
where $D_0 = P a^2$ is the free diffusion coefficient.

\section*{Appendix B}

In this appendix we calculate the mean-square displacement, $\langle (\Delta z)^2 \rangle = 2 D^* t$, from the transition probabilities in order to obtain the single-particle diffusion coefficient $D^*$. As in Appendix A, we do not use subindex $A$ or $B$ to specify the kind of particle, since the resulting equations are the same for both species.

Let us consider a time interval $\Delta t$ small enough so that jumps occur only between neighboring cells. If the tagged particle is in cell $i$ at time 0, the probabilities to have a jump to right or left after a time $\Delta t$ are $W_{i,i+1} \,\Delta t$ and $W_{i,i-1} \,\Delta t$ respectively. Then, the average value of $(\Delta z)^2$ at time $\Delta t$ is
\begin{align}
\langle (\Delta z)^2 \rangle &= a^2 (W_{i,i+1} + W_{i,i-1})  \,\Delta t \nonumber \\
&= a^2 P e^{-\beta(\theta-1)V_i/2}  \nonumber \\
&\times \left(e^{-\beta(\theta+1)V_{i+1}/2} \right.\left. \mbox{}  +  e^{-\beta(\theta+1)V_{i-1}/2} \right)\Delta t.
\end{align}
As in Appendix A, we assume smooth spatial variations and replace $V_i \rightarrow V$, $V_{i+1} \rightarrow V + \dpar{V}{z} a$, $V_{i-1} \rightarrow V - \dpar{V}{z} a$. We obtain, up to order 2 in $a$,
\begin{align}
\langle (\Delta z)^2 \rangle &= a^2 P e^{-\beta(\theta-1)V/2} e^{-\beta(\theta+1)V/2} \nonumber \\
& \ \ \ \times \left\{ \exp\left[-\beta a (\theta+1)\dpart{V}{x}/2\right] \right. \nonumber \\
& \ \ \ \left. \mbox{} + \exp\left[\beta a (\theta+1)\dpart{V}{x}/2\right]   \right\} \Delta t \nonumber \\
&= a^2 P e^{-\beta \theta V}  2  \Delta t.
\end{align}
Then, the single-particle diffusion coefficient is
\begin{equation}
D^* = D_0 \, e^{-\beta \theta V}
\end{equation}
with $D_0 = P a^2$.

\section*{Appendix C}

Here, we show the specific polynomial fits of the experimental data used to represent the functional behavior of $\theta(x)$  and $\beta V (x)$ in our simulations.
For the interpolation parameter $\theta(x)$ we obtained:

\begin{align*}
 \theta_{\text{Al in Al-Cu}} &= 0.061 + 0.275 x - 3.708 x^{2} + 9.108 x^{3}\\ &- 9.199 x^{4} + 3.297x^{5}\\
 \theta_{\text{ Cu in Al-Cu}} &= 1.02 - 1.48 x +0.93 x^{2}\\
 \theta_{\text{ Au in Au-Ag}} &= 3.39 x - 14.85 x^{2} + 34.33 x^{3} \\&- 51.95 x^{4} +41.98 x^{5} - 13.26 x^{6}\\
 \theta_{\text{ Ag in Au-Ag}} &= 5.468 x - 14.183 x^{2} + 14.010 x^{3} - 5.137 x^{4}\\
 \theta_{\text{ Au in Au-Ni}} &= 6.00 - 56.59 x + 285.14 x^{2} - 680.51 x^{3} \\&+ 850.85 x^{4} - 540.14 x^{5} + 137.25 x^{6}\\
 \theta_{\text{ Ni in Au-Ni}} &= 1.00 - 1.60 x - 2.49 x^{2}
\end{align*}

For the mean-field potencial $\beta V(x)$, we got:
\begin{align*}
  \beta V_{\text{Al in Al-Cu}} &= 24.67 x - 35.82 x^{2} + 25.97 x^{3} - 7.91 x^{4}\\
  \beta V_{\text{Cu in Al-Cu}} &= 1.19 x + 4.47 x^{2} - 2.51 x^{4}\\
  \beta V_{\text{Au in Au-Ag}} &= 3.13 x - 1.57 x^{2}\\
  \beta V_{\text{Ag in Au-Ag}} &= 1.92 x - 2.64 x^{2} + 4.84 x^{3} - 2.93 x^{4}\\
  \beta V_{\text{Au in Au-Ni}} &= -6.25 x + 5.42 x^{2} - 1.42 x^{4}\\
  \beta V_{\text{Ni in Au-Ni}} &= -5.10 x + 5.55 x^{2} - 2.43 x^{3}
\end{align*}

\section*{Acknowledgments}
The authors acknowledge H. M\'artin for illuminating discussions. This work was partially supported by Consejo Nacional de Investigaciones Cient\'ificas y T\'ecnicas (CONICET, Argentina, PIP 112 201501 00021 CO).

\bibliography{mybib}{}
\bibliographystyle{ieeetr}
\end{document}